\documentclass[acmsmall]{acmart}

\usepackage{graphicx}
\usepackage{textcomp}
\usepackage{xcolor}
\usepackage{balance}
\usepackage{booktabs}
\usepackage{CJKutf8}
\usepackage{tikz}
\usepackage{forest}
\usepackage{booktabs}
\usepackage{makecell}
\usepackage{float}
\usepackage{algorithm}
\usepackage{algpseudocode}
\usepackage{xspace}
\usepackage{listings}
\usepackage[symbol]{footmisc}
\usepackage{longtable}
\usepackage{nicematrix}
\usepackage{colortbl}
\usepackage{caption}
\usepackage{fontawesome5}
\usepackage{pifont}
\usepackage{subcaption}
\usepackage{multirow}
\usepackage{booktabs}
\usepackage{arydshln}
\usepackage{tabularx}
\usepackage{soul}
\usepackage{hyperref}
\usepackage[]{collab}

\newcolumntype{C}[1]{>{\centering\arraybackslash}p{#1}}

\definecolor{lightgrey}{rgb}{0.96, 0.96, 0.96}
\definecolor{grey}{rgb}{0.92, 0.92, 0.92}
\definecolor{keycolor}{HTML}{2B4983}
\definecolor{valuecolor}{HTML}{61982D}

\newcommand\alias{\textsc{Tracezip}\xspace}
\newcommand\name{Span Retrieval Tree\xspace}
\newcommand\sname{SRT\xspace}

\collabAuthor{zb}{teal}{Zhuangbin}

\AtBeginDocument{%
  }

\begin{document}

\title{\alias: Efficient Distributed Tracing via Trace Compression}

\author{Zhuangbin Chen}
\orcid{0000-0001-5158-6716}
\affiliation{%
  \institution{School of Software Engineering, Sun Yat-sen University}
  \city{Zhuhai}
  \country{China}
}
\email{chenzhb36@mail.sysu.edu.cn}

\author{Junsong Pu}
\orcid{0009-0002-8309-1384}
\affiliation{%
  \institution{Beijing University of Posts and Telecommunication}
  \city{Beijing}
  \country{China}
}
\email{angrychow@bupt.edu.cn}

\author{Zibin Zheng}
\orcid{0000-0002-7878-4330}
\affiliation{%
  \institution{School of Software Engineering, Sun Yat-sen University}
  \city{Zhuhai}
  \country{China}
}
\email{zhzibin@mail.sysu.edu.cn}
\authornote{Zibin Zheng is the corresponding author.}



\renewcommand{\shortauthors}{Zhuangbin Chen, Junsong Pu, and Zibin Zheng}

\begin{abstract}
\textit{Distributed tracing} serves as a fundamental building block in the monitoring and testing of cloud service systems.
To reduce computational and storage overheads, the \textit{de facto} practice is to capture fewer traces via sampling.
However, existing work faces a trade-off between the completeness of tracing and system overhead.
On one hand, \textit{head-based sampling} indiscriminately selects requests to trace when they enter the system, which may miss critical events.
On the other hand, \textit{tail-based sampling} first captures all requests and then selectively persists the edge-case traces, which entails the overheads related to trace collection and ingestion.
Taking a different path, we propose \alias in this paper to enhance the efficiency of distributed tracing via \textit{trace compression}.
Our key insight is that there exists significant redundancy among traces, which results in repetitive transmission of identical data between services and the backend.
We design a new data structure named \name (\sname) that continuously encapsulates such redundancy at the service side and transforms trace spans into a lightweight form.
At the backend, the complete traces can be seamlessly reconstructed by retrieving the common data that are already delivered by previous spans.
\alias includes a series of strategies to optimize the structure of \sname and a differential update mechanism to efficiently synchronize \sname between services and the backend.
Our evaluation on microservices benchmarks, popular cloud service systems, and production trace data demonstrates that \alias can achieve substantial performance gains in trace collection with negligible overhead.
We have implemented \alias inside the OpenTelemetry Collector, making it compatible with existing tracing APIs.
\end{abstract}

\setcopyright{acmlicensed}
\acmJournal{PACMSE}
\acmYear{2025} \acmVolume{2} \acmNumber{ISSTA} \acmArticle{ISSTA019} \acmMonth{7}\acmDOI{10.1145/3728888}

\ccsdesc[500]{Computer systems organization~Cloud computing}
\ccsdesc[500]{Computer systems organization~Reliability}
\ccsdesc[500]{Computer systems organization~Maintainability and maintenance}
\ccsdesc[500]{Computer systems organization~Redundancy}
\ccsdesc[500]{Software and its engineering~Software maintenance tools}

\keywords{Distributed tracing, Trace compression, Cloud computing, System monitoring}


\maketitle

\section{Introduction}

In modern cloud systems, the adoption of loosely coupled designs for applications and services has marked a significant paradigm shift in software architecture.
While such modular design brings the benefits of scalability and operational flexibility, it also complicates different aspects of software development and maintenance, including testing, debugging and diagnosing.
This can be largely attributed to the cascading effect of failures~\cite{DBLP:conf/icse/DangLH19,DBLP:conf/sigsoft/ChenKLZZXZYSXDG20}, i.e., a single failure in one service can quickly propagate to other interconnected services and components.
As such, distributed tracing has emerged as an essential reliability management solution in cloud systems.
This is primarily due to its ability to provide a comprehensive and granular view of the interactions between services, allowing for the precise identification of where and how failures occur and spread.

In cloud service systems, unusual and edge-case system behaviors, such as tail latency and resource contention, are rare by definition.
To achieve high coverage of outlier system events, it is necessary to trace \textit{all} requests.
In production environments, this may result in substantial trace data, incurring significant overhead and costs related to trace generation, collection, and ingestion.
To mitigate this problem, various \textit{trace sampling} techniques have been proposed, which can be categorized into two main types, i.e., \textit{head-based sampling}~\cite{DBLP:conf/nsdi/ZhangXAVM23} and \textit{tail-based sampling}~\cite{DBLP:conf/IEEEcloud/ChenJSLZ24,DBLP:conf/icws/HuangCYCZ21,DBLP:conf/sigsoft/HeFLZ0LR023,DBLP:conf/nsdi/ZhangXAVM23}.
Head-based sampling uniformly collects traces at random based on a small sampling rate (e.g., 1\%~\cite{DBLP:conf/nsdi/ZhangXAVM23,DBLP:conf/cloud/Las-CasasPAM19,DBLP:conf/sosp/KaldorMBGKOOSSV17}).
The sampling decision is made before request execution, and only the sampled requests will be traced.
On the other hand, tail-based sampling captures traces for all requests, and decides whether to retain a trace based on the execution details such as latency and HTTP status code.
In this regard, many machine learning techniques have been applied to automatically select informative and uncommon traces.
However, given the inherent unpredictability of the true value of traces, trace sampling struggles to capture the full spectrum of critical information necessary for effective software testing and failure diagnosis.

In this paper, we propose \alias, an online and scalable solution to address the overhead of distributed tracing by \textit{trace compression}.
\alias transforms spans into a concise representation upon their generation at the service side, which can be seamlessly decompressed at the backend side.
This strategy significantly reduces the volume of data that needs to be transmitted.
We discuss two possible solutions for this end.
The first is \textit{offline log compression}~\cite{DBLP:conf/fast/WeiZWLZCSZ21,DBLP:conf/icse/LiZL024,DBLP:conf/osdi/RodriguesLY21,wang2024muslope}, which condenses log data after it has been aggregated at the backend for long-term persistence.
As the primary goal is to save storage space, this approach often involves sophisticated algorithms which fail to meet the real-time requirements for trace collection.
Moreover, it often requires processing the entire log dataset to achieve the optimal performance.
In our scenario, online learning capability is crucial as traces are continuously generated in a stream.
\textit{General-purpose compression algorithms} (e.g., gzip and bzip) may also seem an out-of-box solution.
However, they are designed for encoding arbitrary binary sequences, which can only exploit redundant information within a short sliding window (e.g., 32KB in gzip's Deflate algorithm).

To pursue more effective compression, \alias harnesses the global redundancy inherent in the structures of trace spans. 
Specifically, we design a new data structure, \textit{\name} (\sname), based on the principles of prefix trees, which is able to continuously extract the set of key-value pairs commonly shared across spans.
The \sname, when synchronized between services and the backend, serves as a reference mechanism to retrieve the identical data that have been previously transmitted by other spans.
To manage computational and space complexity, \alias constantly restructures \sname into its most compact form and employs mapping techniques to further reduce its size.
We also propose a differential update mechanism to effectively synchronize \sname between services and the backend.
\alias is orthogonal to trace sampling methods and can work with log compression techniques once trace data have been efficiently transmitted to the backend.

We have implemented \alias inside OpenTelemetry Collector~\cite{opentelemetry_collector}, one of the most popular tracing frameworks, making it compatible with existing tracing APIs.
We deploy \alias to collect traces for an open-source microservices benchmark (i.e., Train Ticket~\cite{DBLP:journals/tse/ZhouPXSJLD21}) and six application backend components in cloud environments (i.e., gRPC, Apache Kafka, Servlet, MySQL, Redis, MongoDB).
We also evaluate the compression performance of \alias on Alibaba production traces~\cite{DBLP:conf/cloud/LuoXLYXZDH021,DBLP:conf/icpp/WangLWJCWDXHYZ22}.
The experimental results show that \alias offers around 10\%$\sim$45\% performance gain when working in conjunction with traditional compression schemes, i.e., gzip, bzip2, lzma.
Moreover, \alias demonstrates negligible space overhead (i.e., several megabytes) and high efficiency.

The major contributions of this work are as follows:

\begin{itemize}
    \item We propose \alias, the \textit{first} online trace compression system by leveraging the inherent redundancy in trace span data.
    This information is captured by a new data structure we developed, the \name (\sname).
    By sharing \sname between services and the tracing backend, we can eliminate the redundancy associated with the repetitive transmission of identical data across multiple spans.

    \item We have implemented \alias inside OpenTelemetry Collector with a series of optimization strategies for \sname restructuring and synchronization.
    Experiments on both open-source systems and production trace dataset demonstrate that \alias can effectively compress traces with MB-scale space overhead and superior efficiency.
    The implementation and data of \alias are publicly available\footnote{\url{https://github.com/OpsPAI/TraceZip}}.
\end{itemize}

The remainder of the paper is organized as follows.
Section~\ref{sec:background} introduces the background of distributed tracing and the motivation of this work.
Section~\ref{sec:methodology} and~\ref{sec:implementation} describes the proposed methodology and system implementation.
Section~\ref{sec:exp_eval} presents the experiments and results. 
Section~\ref{sec:related_work} discusses the related work.
Finally, Section~\ref{sec:conclusion} concludes this work.



\section{Background and Motivation}
\label{sec:background}

\subsection{Distributed Tracing}

Distributed tracing provides a detailed end-to-end view of requests as they traverse interconnected and multi-tier cloud service systems.
The building blocks of a trace are called \textit{spans}.
They represent the individual slices of work performed across different machines and components that are visited by the request.
Each span encapsulates various attributes, including \textit{span name}, \textit{parent span ID}, \textit{span ID}, \textit{start/end timestamps}, \textit{events}, etc.
This information enables the correlation and analysis of the request's lifecycle.

\begin{figure}
    \centering
    \includegraphics[width=0.5\linewidth]{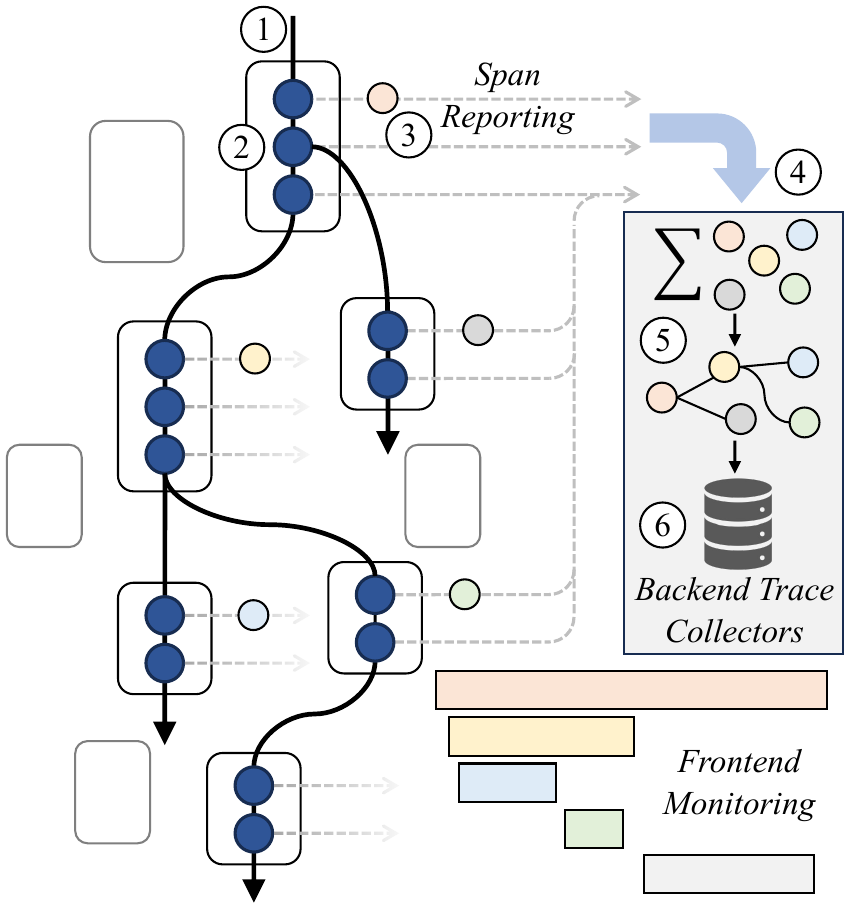}
    \caption{A Typical Procedure of Distributed Tracing}
    \label{fig:distributed_tracing}
\end{figure}

Figure~\ref{fig:distributed_tracing} illustrates a typical procedure of tracking requests~\cite{DBLP:conf/nsdi/ZhangXAVM23,opentelemetry} in modern distributed tracing frameworks, such as Jaeger~\cite{jaeger} and Zipkin~\cite{zipkin}.
Upon the arrival of a new request to the application, the tracing framework assigns it a unique \texttt{trace\_id} (\ding{192}).
However, not all requests will actually be traced, which is indicated by a flag, \texttt{sampled}.
Both \texttt{trace\_id} and \texttt{sampled} will then be propagated along the request at the application level, which is important for the completeness and coherence of a trace.
If \texttt{sampled} is set, each component (e.g., a microservice instance) that handles the request will generate trace data (\ding{193}), i.e., a span, using the tracing framework's client library (e.g., OpenTelemetry).
The places that emit spans are called a \textit{tracepoint}, and there could be multiple tracepoints serving different requests or different operations within a single request.
The framework's client library then enqueues, serializes, and transmits trace spans (\ding{194}) to its centralized backend collection infrastructure, or simply backend.
The backend is responsible for continuously receiving (\ding{195}), processing (\ding{196}), and storing (\ding{197}) trace data.
Based on the \texttt{trace\_id}, \texttt{parent\_span\_id}, and \texttt{span\_id}, the backend can assemble spans that were dispersed across different components into a single coherent trace.

Given the details provided by these spans, operators can closely monitor and understand the impact that one component may have on the others.
This makes trace data particularly useful for troubleshooting cross-component problems in large distributed systems.
However, traces can be produced at high volume, incurring significant network, compute, and storage costs.
In production scenarios, Google is estimated to generate approximately 1,000 TB of raw traces on a daily basis.
Netflix needs to manage more than 2 billions of daily requests.
To mitigate overheads, existing tracing frameworks often apply head-based sampling to trace only a small fraction of requests by randomly setting the \texttt{sampled} flag (\ding{192}).
This will inevitably increases the risk of overlooking system edge-case behaviors.
On the other hand, tail-based sampling utilizes a filtering strategy, i.e., only persisting traces that exhibit outliers symptoms (\ding{197}), e.g., high tail latency, error codes.
However, tail-based sampling entails enormous costs, as it must trace all requests and ingest the trace data to make sampling decisions.

In this work, we take an orthogonal path to address the overhead challenges.
We alleviate the lossy nature of sampling-based schemes by tracing more requests, yet without increasing the transmission overhead.
Such a design can enhance the monitoring capability of edge-case behaviors.
Our core idea lies in the observation that there exists a large amount of redundancy in trace data.
Specifically, for each tracepoint, the generated spans may have repeated attribute values and events.
This renders repetitive transmissions of identical trace data, and we see this could be an opportunity to improve the tracing efficiency.
Recognizing this, we conduct a study to examine the redundancy inherent in trace data.
The insights obtained serve as essential design principles of our approach.


\subsection{A Study of the Redundancy in Trace Data}
\label{sec:redundancy_study}


In this section, we present our study regarding the \textit{redundancy} of trace data.
By redundancy, we mean the recurrence of identical information, e.g., attributes and events, that is observed repeatedly across multiple traces.
The identification of such repetitive patterns presents an opportunity to enhance the efficiency of distributed tracing through the application of compression techniques.
By strategically reducing the redundancy, we can optimize the transmission overhead of trace data, thereby improving the scalability of the tracing infrastructure without compromising data integrity.

The studied systems include an open-source microservices benchmark, i.e., Train Ticket~\cite{DBLP:journals/tse/ZhouPXSJLD21}, which is popular in cloud-related research fields, and six application backend components that are selected for their widespread use in modern cloud systems, i.e., gRPC (Client and Server), Apache Kafka (Producer and Consumer), Servlet, MySQL, Redis, and MongoDB.
They cover a diverse functionalities including message queuing, HTTP communication, remote procedure calls, database management, etc.
To collect their traces, we leverage the zero-code instrumentation capabiltiy of OpenTelemetry~\cite{opentelemetry_zero_code}.
It allows the collection of observability data, i.e., logs, metrics, and traces, for applications without the need to modify the source code.
This is achieved by using libraries, plugins, or agent to instrument the libraries used by applications.
It supports many programming languages (e.g., Java, Go, Python) and a wide range of popular libraries and frameworks, including requests and responses, database calls, message queue calls.
By injecting typical workloads to the studied systems (Section~\ref{sec:deployed_cloud_systems}), we collect more than 40GB of trace data in total.

\begin{figure}
    \centering
    \includegraphics[width=0.8\linewidth]{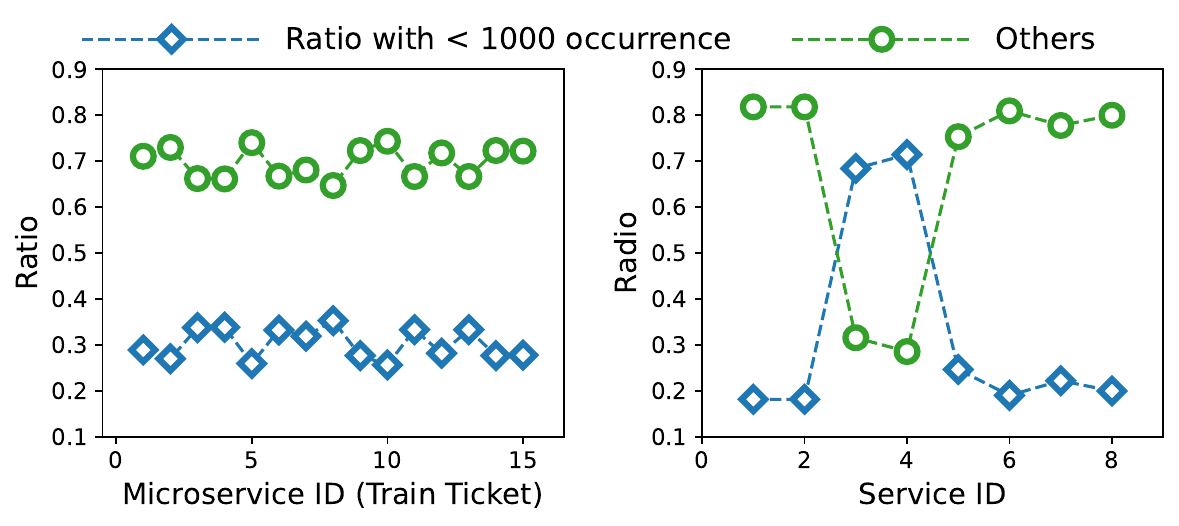}
    \caption{Trace Data Redundancy Analysis}
    \label{fig:trace_redundancy}
\end{figure}

We quantify trace redundancy by measuring the proportion of duplicate key-value (KV) pairs generated by each service.
As a KV pair represents the fundamental unit of information within a trace, assessing its repetition enables us to gauge the degree of information overlap.
Specifically, for each KV pair, we first count the number of its occurrence and then calculate its ratio over the total number of KV pairs in the dataset.
For example, in a trace dataset containing 100 KV pairs, if two specific pairs appear once and ten times respectively, their redundancy ratios would be 1\% and 10\%, respectively.
We calculate the fraction of KV pairs within each service which occur less than 1,000 times, as well as those exceeding this threshold, as shown in Figure~\ref{fig:trace_redundancy}.
We make the following two important observations.


\textbf{Traces are highly redundant}.
Based on the figures, we can see that for different microservices in Train Ticket, around 70\% of the total KV pairs are highly repetitive.
Similar situations can be found in the application components.
The only exception is Apache Kafka, whose KV pairs tend to be unique.
The reason behind is that Kafka's traces include the data from its message queues, increasing the randomness of its KV pairs.
We will discuss its impact on compression during system evaluation (Section~\ref{sec:exp_eval}).
The results indicate that a significant portion of the trace data is characterized by redundant information.
An important reason is that services often engage in standard interactions and perform routine operations that generate trace data with similar patterns.
By capturing the redundancy upon the generation of spans at the service side (\ding{192}), we can preemptively eliminate the transmission of repetitive data that already exist in the backend.
The backend can easily reconstruct the complete spans based on their redundancy patterns (\ding{197}).

\textbf{There exists structural redundancy among attributes}.
We also observe that there exists certain redundancy at span level.
OpenTelemetry's semantic conventions~\cite{opentelemetry_semantic_conventions} offer standardized guidelines for naming common attributes across different kinds of operations and data.
This is essential for maintaining the uniformity and compatibility of naming scheme across languages, libraries, and platforms.
In the naming scheme, attribute keys are organized into hierarchical namespaces to indicate their context or category, e.g., \texttt{network.local.address}, \texttt{network.local.port}, \texttt{network.peer.address}.
We can see that they share some common words.
The attribute values also exhibit similar commonality, e.g., Java Exception has \texttt{java.io.IIOException}, \texttt{java.io.EOFException}.
By eliminating such fine-grained redundancies, we can further reduce the overhead associated with span transmission.
\section{Methodology}
\label{sec:methodology}

\subsection{Overview}

\begin{figure}[t]
    \centering
    \includegraphics[width=0.62\linewidth]{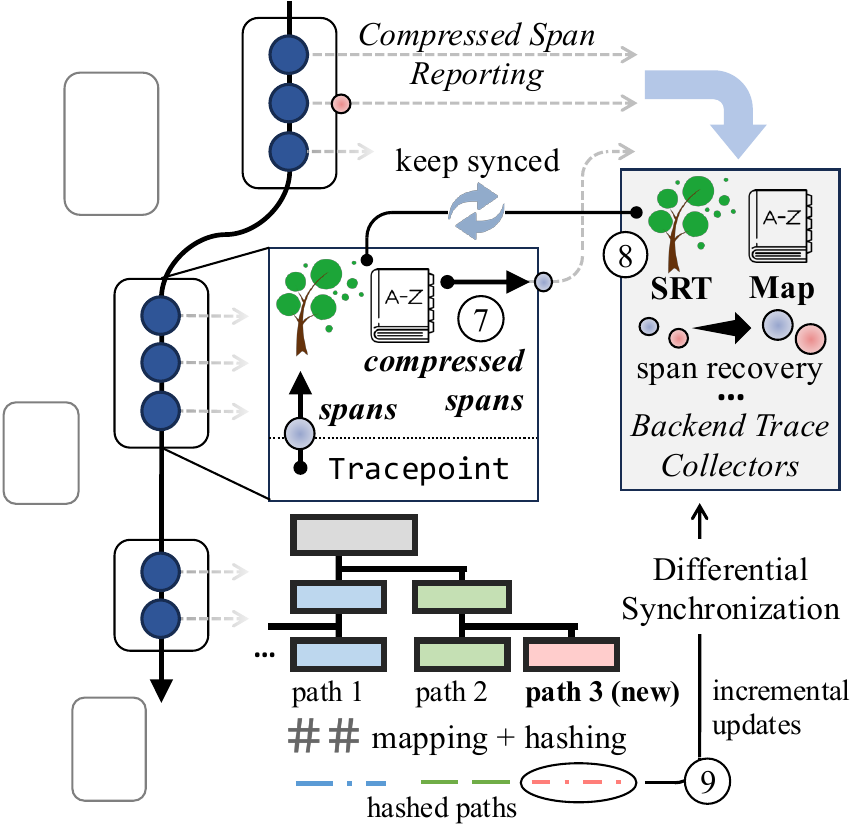}
    \caption{System Architecture of \alias}
    \label{fig:tracezip_system}
\end{figure}

In this section, we present the design of \alias.
The system architecture is illustrated in Figure~\ref{fig:tracezip_system}, in which we add a \textit{compression module} (\ding{198}) to the service side and a corresponding \textit{decompression module} (\ding{199}) to the entry of the backend trace collectors.
In the compression module, we maintain two data structures, namely, a \name (\sname) and a dictionary, to constantly capture the redundancy across the spans.
Upon the generation of a span at the tracepoint, it undergoes compression utilizing the above data structures.
If the span carries a new redundancy pattern, it will be seamlessly integrated into the \sname and dictionary.
This integration is crucial as it enriches the structures, thereby facilitating the compression for subsequent spans.
We accelerate the above process by employing a combination of mapping and hashing techniques.
At the decompression module, the spans are restored to their original form by referring to the SRT and the dictionary.
To ensure a consistent and reliable data transmission,  it is imperative that these data structures are accurately synchronized between the service and backend sides.
To achieve this, we develop a differential update mechanism (\ding{200}).
This mechanism is designed to precisely pinpoint and propagate only the incremental changes in the data structures, ensuring an efficient synchronization process that minimizes overhead while maximizing data consistency.

\subsection{Span Format Conventions}

To compress spans by leveraging their recurring patterns, we first stipulate the format of a span.
For simplicity and readibility, we assume that a span adheres to the standard JSON data format that defines it as a structured set of key-value pairs.
The key is a string, while the value can be either primitive types (strings, numbers, booleans, and null) or two structured types (nested key-value pairs and arrays).
This aligns with the format specifications used in many tracing frameworks and tools, e.g., OpenTelemetry~\cite{opentelemetry_traces}, Jaeger~\cite{jaeger}, Zipkin~\cite{zipkin}.
Typical fields (keys) of a span include: \textit{Name} (a human-readable string representing the operation done), \textit{Parent Span ID} (the span that caused the creation of this span, empty for root spans), \textit{Start and End Timestamps} (the start and end time of the span), \textit{Span Context} (the context of the span including the trace ID, the span ID, etc.), \textit{Attributes} (key-value pairs representing additional information about the span), \textit{Span Events} (structured log messages/annotations on a span), etc.
It is important to note that our proposed algorithm is not restricted to JSON or any particular serialization format.
For example, \alias can work effectively when Protobuf (Protocol Buffers)~\cite{protobuf} is used for trace data serialization.
With Protobuf's powerful deserialization capabilities, we can leverage its reflection-like APIs or direct-access methods to dynamically access the fields and values of spans.
Additionally, Protobuf is designed to be backward and forward compatible, allowing us to modify the message definition by adding or removing fields while maintaining compatibility with older data.
Such operations are essential for reducing trace redundancy, e.g., removing span elements that are deemed repetitive.

We also assume that spans possess \textit{structural locality}, meaning that during the continuous execution of a service or component, all spans sharing a common span Name will exhibit an identical structure.
In other words, spans with the same Name will consistently retain the same set of keys (e.g., attributes, tags, and metadata), differing only in the specific values associated with them.
This assumption arises naturally from the way distributed tracing systems operate, where spans typically represent predefined operations or events within the service workflow.
These operations are implemented as part of the service's codebase, which enforces a fixed schema or structure for spans generated by specific instrumentation points.
This structural consistency allows for reliable trace analysis, optimization, and redundancy reduction, as the predictability of span structures minimizes the need for per-span schema discovery during processing.




\subsection{Span Retrieval Compression and Uncompression}
\label{sec:compression_uncompression}

A straightforward approach to compressing spans involves the use of a dictionary.
This method creates a dictionary where every unique key and value is assigned a unique identifier.
During the compression process, the keys and values of each span are substituted by the corresponding identifiers.
The size of the span can then be reduced as the identifiers are much smaller than the original data.
However, as revealed by our empirical study, there can still be redundant information among the identifiers.
The pure dictionary approach compresses data on a one-to-one basis, i.e., one identifier corresponds to one KV pair.
If multiple spans share a collection of common key-value pairs, it is possible to utilize a single identifier to represent this entire set of shared pairs, thereby amplifying the compression efficiency.
Thus, we propose to leverage the correlations among the values of spans to further eliminate repetitive information.






\begin{algorithm}
    \caption{Span Retrieval Tree (SRT) Reconstruction and Span Compression}
    \label{algo:srt}
    \begin{algorithmic}[1]
    \State \textbf{Input:} a stream of continuously generated \textit{spans}, a threshold $\psi$
    \State \textbf{Output:} a constructed SRT $\mathcal{T}$, \textit{compressed spans}
    \State Initialize an empty SRT $\mathcal{T}$
    \For{each \textit{span} in \textit{spans}}
        \If{\textit{span Name} not in $\mathcal{T}$}
            \State Chain all key-value pairs of \textit{span} and add the path to the root of $\mathcal{T}$
            \State Assign an identifier to this new path
        \Else
        \For{each \textit{key} at every depth of $\mathcal{T}$} \Comment{traverse $\mathcal{T}$ from the root to the leaf}
            \State Get the corresponding \{\textit{key}: \textit{value}\} from \textit{span}
            \If{\{\textit{key}: \textit{value}\} \textbf{does not exist} at the current depth of $\mathcal{T}$}
                \State Chain the remaining key-value pairs of \textit{span} and extend a new branch from the direct parent node of \textit{key}
                \State Calculate the number of unique nodes at each depth of $\mathcal{T}$
                \State Move the keys to the leaf whose unique value number exceeds $\psi$, i.e., \textit{local fields}
                \State Reorder the keys of $\mathcal{T}$ based on the ascending number of their unique values
                \State Reassign path identifiers
                \State \textbf{break}
            \EndIf
        \EndFor
        \EndIf

    \State Compress \textit{span} based on the corresponding path identifier and the values of local fields
    \EndFor
    \end{algorithmic}
\end{algorithm}

\begin{figure}[t]
    \centering
    \includegraphics[width=0.74\linewidth]{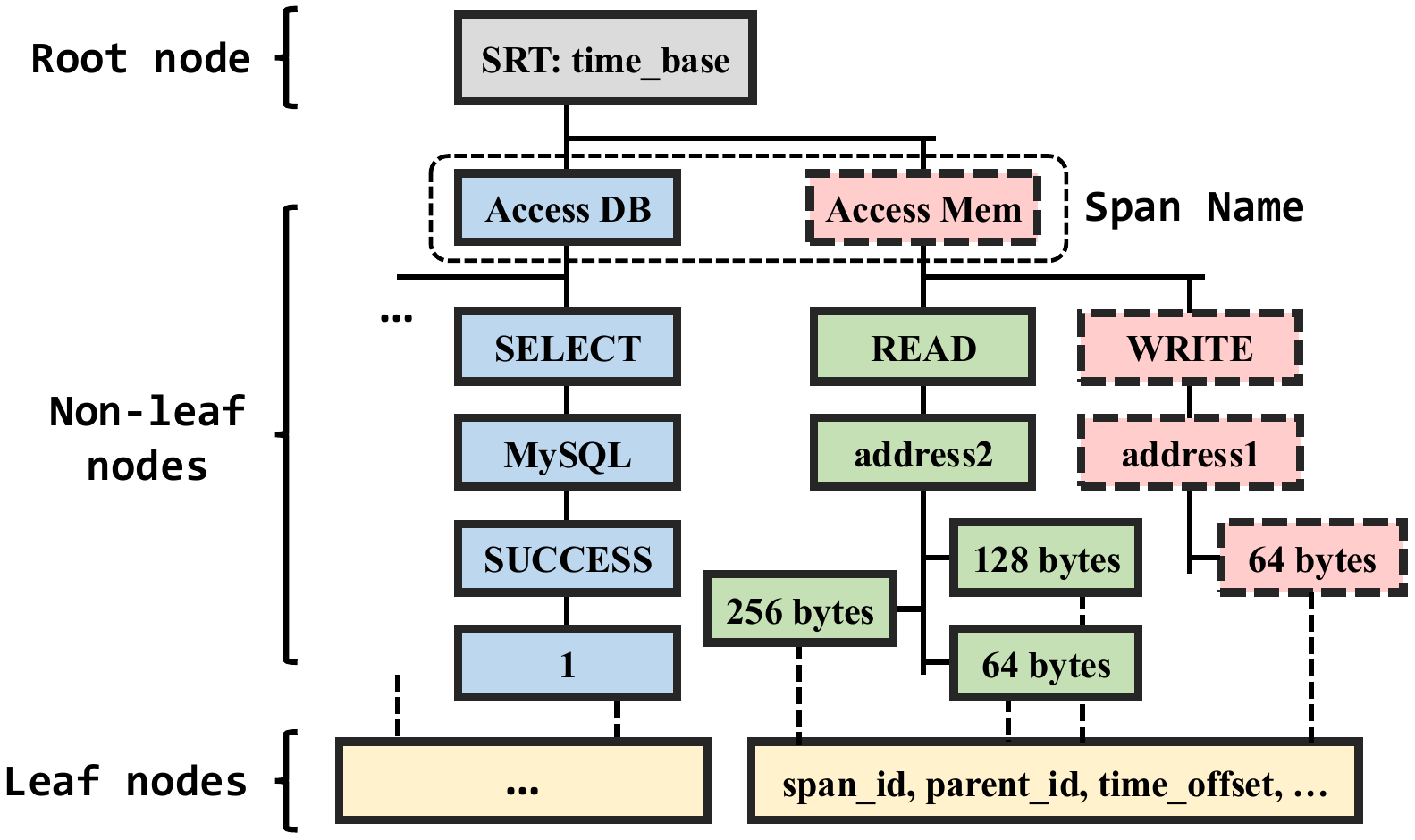}
    \caption{An Example of \name}
    \label{fig:span_retrieval_tree}
\end{figure}

Our idea is that for spans generated in each service instance, we organize their key-value pairs as a prefix-tree-like data structure, i.e., \sname.
The \sname functions as a multi-way tree, with all non-leaf nodes (except for the root) associated with a key-value pair.
For each type of span (which is identified by a unique span Name), there is only one leaf node connected to all the last non-leaf nodes stemming from it.
This leaf node holds a collection of keys without values.
Figure~\ref{fig:span_retrieval_tree} illustrates an example of SRT, where the gray node and the yellow nodes represent the root and the leaves, respectively, while the remaining are the non-leaf nodes.
Each span can be ``spelled out'' by tracing a path from the root down to the leaf.
The path of \sname represents the set of KV pairs shared across multiple spans.
The non-leaf nodes contain the fields that are more repetitive, which we refer to as \textit{universal fields}.
Although spans may exhibit commonality, they will still have some unique KV pairs, such as those related to ID and timestamps.
We refer to such pairs as \textit{local fields} and only store their keys at the leaf.
The rationale is that such unique fields are incompressible, i.e., not shared with other spans, so we discard their values.
\textit{Based on \sname, a span can be represented as a unique path identifier plus its exclusive values that are extracted according to the keys in the leaf node.}
Each path identifier collectively represents the KV pairs shared among spans, instead of one identifier for each key and value.
Since these common KV pairs constitute a significant portion, the trace size can be substantially reduced, enhancing the overall efficiency.


\begin{table}
    \centering
    \caption{Span Examples of a Data-processing Service}
    \label{tab:span_examples}
    \begin{subtable}{1\textwidth}
    \small
        \centering
        \begin{NiceTabular}{c|c|c|c|c|c}
            \specialrule{0.35mm}{0em}{0em}
            \rowcolor{grey}\textbf{name} & \textbf{operation} & \textbf{address} & \textbf{data\_size} & \textbf{span\_id} & \textbf{others}\\
            \specialrule{0.15mm}{0em}{0em}
            \specialrule{0.15mm}{.1em}{0em}
            Access Mem & WRITE & address1 & 64 bytes & id1 & ...\\
            Access Mem & READ & address2 & 128 bytes & id2 & ...\\
            Access Mem & READ & address2 & 64 bytes & id3 & ...\\
            Access Mem & WRITE & address1 & 64 bytes & id4 & ...\\
            Access Mem & READ & address2 & 256 bytes & id5 & ...\\
            \specialrule{0.35mm}{0em}{0em}
        \end{NiceTabular}
        \caption{Span examples of ``Access Mem''}
    \end{subtable}
    \vfill
    \begin{subtable}{1\textwidth}
    \small
        \centering
        \begin{NiceTabular}{c|c|c|c|c|c}
            \specialrule{0.35mm}{0em}{0em}
            \rowcolor{grey}\textbf{name} & \textbf{type} & \textbf{DB system} & \textbf{status} & \textbf{row.num} & \textbf{others}\\
            \specialrule{0.15mm}{0em}{0em}
            \specialrule{0.15mm}{.1em}{0em}
            Access DB & INSERT & MySQL & SUCCESS & 1 & ...\\
            Access DB & SELECT & MySQL & SUCCESS & 1 & ...\\
            Access DB & DELETE & MySQL & SUCCESS & 1 & ...\\
            \specialrule{0.35mm}{0em}{0em}
        \end{NiceTabular}
        \caption{Span examples of ``Access DB''}
    \end{subtable}
\end{table}

We present our algorithm for \sname construction and span compression (i.e., Algorithm~\ref{algo:srt}) and explain it using span examples in Table~\ref{tab:span_examples}.
Suppose these spans are continuously generated by different tracepoints of a data-accessing service, including memory and database.
Each tracepoint produces a specific type of span with varying attributes.
The algorithm takes the stream of spans as input, and the resulting \sname is shown in Figure~\ref{fig:span_retrieval_tree}.
For each new type of span with a previously unseen span Name, we simply chain all fields of the span (line 6) and add the resulting path to the \sname root.
For example, the first row of Table~\ref{tab:span_examples}-(a) will be structured as \textit{Access Mem}$\hookrightarrow$ \textit{WRITE}$\hookrightarrow$\textit{address1}$\hookrightarrow$\textit{64~bytes}$\hookrightarrow$\textit{id1} (we omit the keys of the nodes and the other attributes), shown as the pink dashed rectangles.
For spans with a known Name, we traverse the \sname from the root to the leaf, and use the key at each depth to retrieve the corresponding key-value pair from the span (line 10).
If a retrieved pair does not exist in the SRT, the remaining key-value pairs are chained to construct a sub-path, which is then added as a new branch to the direct parent node (line 12).
For example, the second row adds a new path, \textit{READ}$\hookrightarrow$\textit{address2}$\hookrightarrow$\textit{128~bytes}$\hookrightarrow$\textit{id2}, to node \textit{Access~Mem}.
Each path of \sname will be assigned a unique identifier, as described in Section~\ref{sec:hashing_acceleration}.


In \sname, once a new path emerges, we calculate the number of distinct nodes at the same depth (line 13), which represents the number of different values of a key, e.g., the key \textit{span.id} has five distinct values \textit{id1}$\backsim$\textit{id5}.
We set a threshold $\psi$ for the size of values a key can have.
A key with too many values will be regarded as a local field and moved to the leaf (line 14).
For example, \textit{span.id} will be in the leaf if $\psi=3$.
With the constructed SRT, the fourth row can be compactly represented as the identifier of the first path, i.e., the pink path, coupled with its unique value, i.e., \textit{id4} (line 21).
Time-related fields such as span start/end time is also a typical local field.
Since spans generated in a short time period will have close temporal fields, we set a \textit{time\_base} at the root node, which allows the leaves to store only the offset relative to the time base.
This is a common way to compress temporal data.
Another special local field is the nested JSON object, such as \textit{\{``attributes'': \{``ip'': ``172.17.0.1'', ``port'': 26040\}\}}.
We represent the nested structure by prefixing the keys of the child JSON object with the parent's key (e.g., \textit{``attributes-ip'': ``172.17.0.1''} and \textit{``attributes-port'': 26040}), which allows the backend to easily restore the original hierarchy.
Technically, spans can extend to any depth as required by the tracing needs.
For the consideration of \sname's size, we set a depth limit, which defaults to two, and convert the values of the overly deep keys into pure string type.
Similar to other string fields, they will be moved to the leaves if exhibiting too much diversity.

After compression, the span data that needs to be transmitted to the backend trace collector become significantly smaller in size, i.e., only the path identifier and the values of local fields specified by the leaf.
At the backend side, the uncompression process to restore the original span is straightforward and efficient.
This involves reconstructing the local fields based on the corresponding values received and combining them with the universal fields based on the path identifier.
In this process, the backend side should keep the latest copy of the \sname and the value of \textit{time\_base}.
We introduce an efficient synchronization mechanism later in Section~\ref{sec:differential_sync}.
For \textit{time\_base}, we periodically reset it, e.g., every second, ensuring that the time offset remains consistently small.

\subsection{Optimizations for \name}

So far, we have introduced the algorithms for span compression and uncompression.
It can be seen that \alias has a small computational complexity.
This is because for each span, these processes involve only a single path traversal of the \sname from the root to a leaf.
However, the issue of space complexity presents a more significant challenge.
The \sname can potentially grow too large and consume an excessive amount of memory.
Besides setting a hard constraint on the memory, we have also identified some opportunities to optimize its size.

\subsubsection{\name Restructuring}

\begin{figure}
    \centering
    \includegraphics[width=0.66\linewidth]{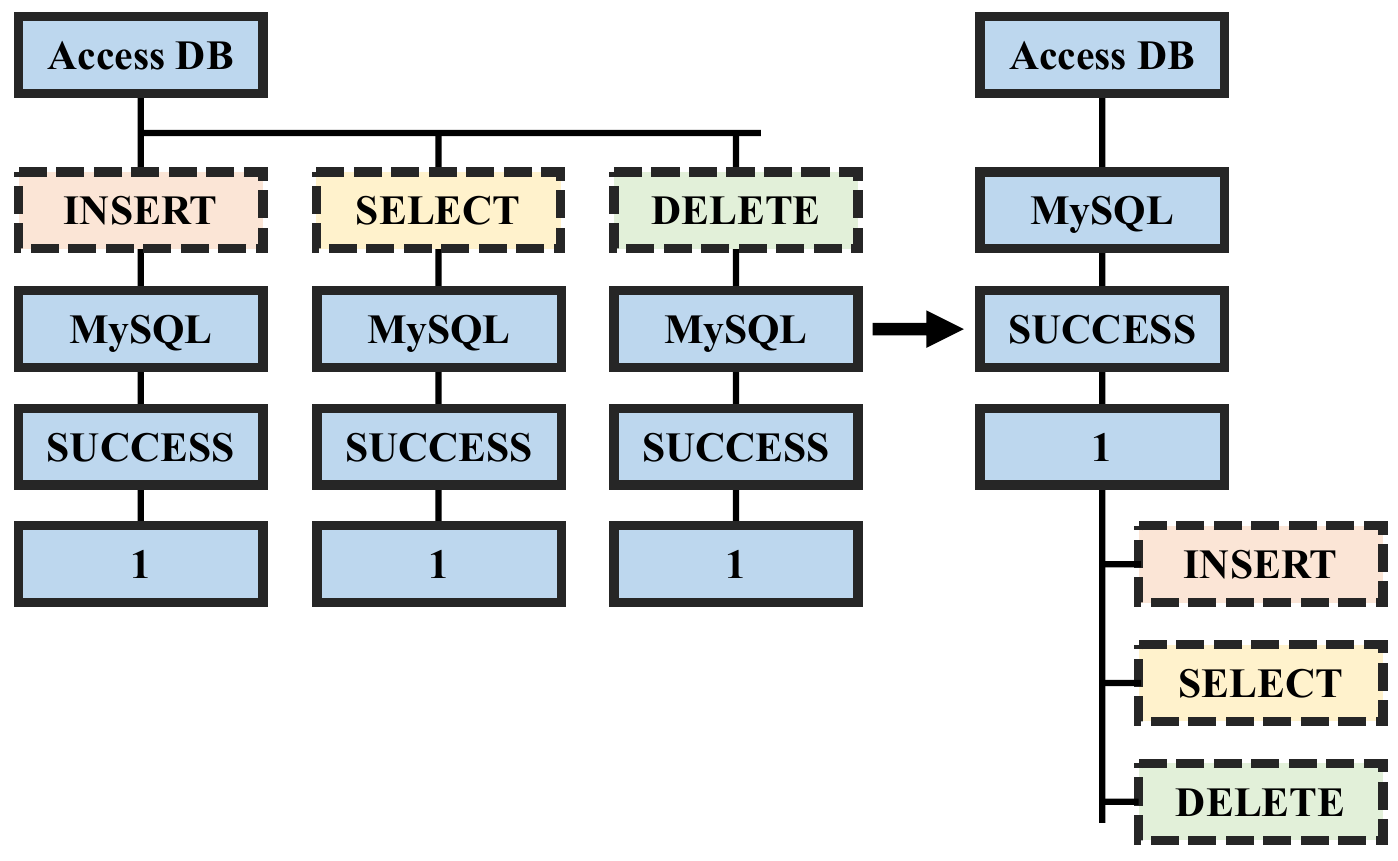}
    \caption{Span Retrieval Tree Restructuring}
    \label{fig:sft_restructuring}
\end{figure}

During the construction of \sname in Figure~\ref{fig:span_retrieval_tree}, we simply follow the left-to-right order of keys in Table~\ref{tab:span_examples} to form the parent-child relations among nodes.
For example, key \textit{address} is the child of \textit{name} and also the parent of \textit{data.size}.
We observe that this may result in a sub-optimal \sname structure.
Specifically, for the \sname in Figure~\ref{fig:sft_restructuring} which is built based on the spans in Table~\ref{tab:span_examples}-(b), we can see that the three paths differ only in the \textit{type} field.
A better structure can be obtained by moving \textit{type} down to the bottom, which avoids the recurrence of the other three fields.
Based on this finding, we propose the following way to restructure the \sname.
In Section~\ref{sec:compression_uncompression}, we have calculated the number of possible values associated with each key once a new path emerges.
If a parent field has more values than its child, we swap their positions in the \sname.
That is, we reorder the keys of \sname based on the ascending number of their unique values (line 15).
After reordering, the identical nodes at the same depth will be merged, e.g., \textit{MySQL}, \textit{SUCCESS}, and \textit{1} in Figure~\ref{fig:sft_restructuring}.
Finally, the path identifiers of the restructured \sname will be adjusted (line 16).



\subsubsection{Mapping-based Tree Compression}
\label{sec:mapping-based tree compression}

Although we have restructured the \sname to eliminate redundant nodes, there could still be repeated keys and values in it.
For example, in Figure~\ref{fig:span_retrieval_tree}, key \textit{data.size} appears in all \textit{data.size} nodes, e.g., \textit{\{``data.size'': ``64 bytes''\}} and two of them also share value \textit{64 bytes.}
Thus, to further compress the size of \sname, we employ a dictionary to map keys/values that occur multiple times to shorter identifiers.
We construct the identifiers using the standard alphanumeric set, i.e., [0-9a-zA-Z].
Initially, the hashed output consists of a single character, from '0' to '9,' followed by 'a' through 'z,' and finally 'A' through 'Z.'
Upon exhausting the single character possibilities, the function increases the length of the hash output to two characters, starting from '00,' '01,' and so forth.
Since each universal field has limited distinct values, i.e., smaller than $\psi$, the dictionary will also be small in size.
Note we do not encode the values of local fields (not in the \sname), which may inevitably make the dictionary too big given their diversity.
Similar to the \sname synchronization process between services and the tracing backend, the dictionary will be sent to the backend every time it undergoes an update.

Based on our empirical study (Section~\ref{sec:redundancy_study}), there exists structural redundancy among the attributes of a span.
We address this issue by examining the ingredients of the attributes.
Specifically, when constructing the dictionary, we encode the common sub-fields shared among spans (instead of the entire fields) as identifiers.
These sub-field identifiers are then used to compose the complete attributes.
Take a span example from OpenTelemetry~\cite{opentelemetry_traces}, which contains the following fields:

\begin{figure}[h]
    \centering
    \includegraphics[width=0.6\linewidth]{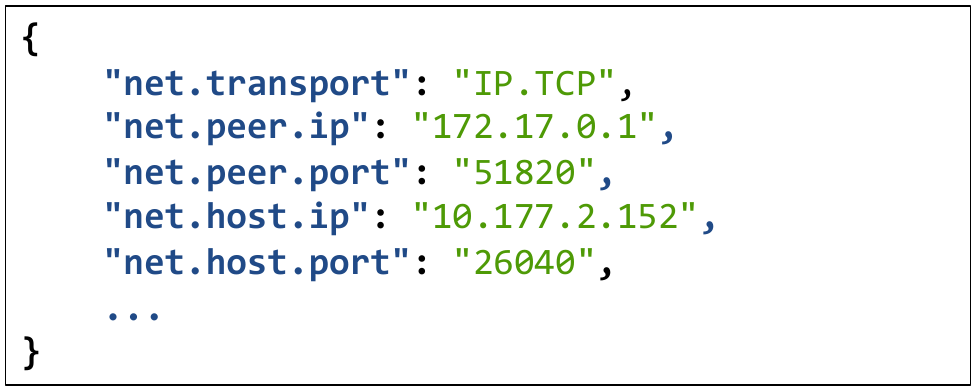}
    \label{fig:structural_correlation}
\end{figure}


\noindent We can see that in the keys, there are some words that appear multiple times, e.g., \texttt{net}, \texttt{host}, \texttt{port}.
Without considering such correlations, we could potentially introduce too much lengthy keys to the dictionary.
To remove such redundancy, we first separate each key into a list of tokens based on delimiters dot (``.'') and underline (``\_''), which are configurable.
When encoding the key, each of its tokens will be mapped to the corresponding identifier.
As the value part exhibits more diversity, we only apply this technique to the keys to avoid too much computational overhead.

\section{Implementation}
\label{sec:implementation}

We have implemented \alias inside the OpenTelemetry Collector with around 3K lines of Golang code.
The OpenTelemetry Collector offers a vendor-agnostic implementation of how to manage telemetry data, which mainly includes four types of components: \textit{exporters}, \textit{processors}, \textit{receivers}, and \textit{extensions}.
We implement the span retrieval compression and decompression on the exporter and receiver, which run at the service side and backend side, respectively.
The exporter is responsible for building and updating the \sname and dictionary, compressing spans on the fly, and sending them to the remote backend.
After accepting the compressed data, the receiver performs span uncompression.
We outline some important details concerning the implementation.

\subsection{Search Acceleration by Hashing}
\label{sec:hashing_acceleration}

A straightforward data structure to implement \sname would be linked representation, which enjoys the benefits of dynamic size and efficient alterations (e.g., insertion and deletion).
However, in linked representation, the tree nodes are not stored contiguously or nearby in memory, potentially leading to more cache misses.
This factor can significantly impede the speed of path search within \sname.
To accelerate the search process, we apply hashing to convert each unique path of SRT to a path identifier, which is similar to that in Section~\ref{sec:mapping-based tree compression}.
Specifically, for each path, starting from the root we join the values of non-leaf nodes sequentially with a comma separator (similar to the CSV format). 
Based on the composed path string, we maintain a \{\textit{path}: \textit{identifier}\} mapping at the exporter.
When a new span is generated at the exporter, we extract the values of its universal fields based on the order in \sname.
The path search can then be quickly done for the span by checking if its path string exists in the map.
We use the \texttt{map} data type in Golang, which provides a highly efficient way to achieve this.
For any updates to the \sname, we only need to renew the affected paths as discussed in the next subsection.

\subsection{Differential Data Synchronization}
\label{sec:differential_sync}

To ensure reliable span compression and uncompression, the exporter and receiver must maintain consistent copies of both the \sname and dictionary structures.
One simple strategy is for the exporter to send the latest versions of these structures upon any update.
However, given that updates often affect only a small segment of the overall structures, sending redundant (i.e., unchanged) data with each update would incur network overhead and potentially defer the uncompression process.
Thus, we implement a differential update mechanism for more resource-efficient synchronization.
The core idea is that at the receiver, instead of maintaining another \sname, we keep a path hashing in the opposite direction, i.e., \{\textit{identifier}: \textit{path}\}.
For any updates to the non-leaf nodes, we can easily pinpoint the affected paths and perform the renewal.
For example, in Figure~\ref{fig:tracezip_system}, the emergence of a new value (denoted by the pink dashed rectangle) gives rise to a novel path, i.e., \textit{path 3}.
In this case, we can add a new entry to the \{\textit{path}: \textit{identifier}\} mapping at the exporter and sync it with the receiver.
For path deletion, the exporter can simply send the corresponding identifier to the receiver for record elimination.
Other updates are essentially a combination of path addition and deletion.

For local fields and the mapping dictionary, it suffices to communicate only the changes to the receiver.
To ensure that the structures at the receiver is not outdated during the transmission of spans, we leverage the batch processor of OpenTelemetry Collector.
It caches the spans sent by SDK until the batch memory is full or its timer expires, instead of immediately forwarding them.
After compressing the spans in the buffer, we will make sure that the SRT and dictionary with updates (if any) have been synced with the receiver side before releasing the data.

\section{Experimental Evaluation}
\label{sec:exp_eval}

In this section, we present the evaluation of \alias.
We first introduce the experimental settings, including the deployed cloud services, the metric for evaluation, and the baseline methods.
Next, we demonstrate the experimental results, which include the effectiveness of trace compression and the analysis of both efficiency and overhead.


\subsection{Experimental Setup}

\subsubsection{Deployed Cloud Systems}
\label{sec:deployed_cloud_systems}

To evaluate the compression performance in a realistic environment, we deploy popular cloud systems and collect their traces using the OpenTelemetry Collector instrumented with \alias.
We serialize the trace data into JSON format and transmit them utilizing HTTP protocols.
The selected services include one microservices benchmark named Train Ticket~\cite{DBLP:journals/tse/ZhouPXSJLD21} and six open-source application components, including gRPC, Apache Kafka, Servlet, MySQL, Redis, MongoDB.

Train Ticket is a railway ticketing application comprising 41 microservices, each responsible for a specific function, such as user authentication, ticket booking, payment processing, and notification.
This benchmark is implemented in different programming languages such as Java, Go, Node.js, Python, etc.
Train Ticket allows a comprehensive evaluation in a multifunctional scenario, which has been widely used in many trace-related topics, including trace sampling~\cite{DBLP:conf/IEEEcloud/ChenJSLZ24}, root cause localization~\cite{DBLP:journals/tse/ZhouPXSJLD21,DBLP:conf/issre/ZhouZPYLLZZD23}, service architecture measurement~\cite{DBLP:conf/sigsoft/0001ZZIGC22}, etc.
In order to replicate a live production environment, we employ Locust~\cite{locust}, an open-loop asynchronous workload generator, to drive the services.
The workloads are directly borrowed from the original work~\cite{DBLP:journals/tse/ZhouPXSJLD21} that introduces the Train Ticket microservices. 

The selected six application components have widespread adoption and play critical roles in modern cloud service architectures.
They represent a diverse cross-section of the technology stack, which play a foundational role in constructing robust, scalable, and high-performance distributed systems.
We generate workloads that reflect real-world usage patterns common in cloud-native and microservices environments.
For communication protocols like gRPC and web service frameworks such as Apache HTTP, we simulate typical traffic and user interactions.
In messaging systems like Kafka, workloads involve data streaming and message processing, while for data storage solutions like MySQL, Redis, and MongoDB, we focus on common database operations such as read/write transactions.
This approach ensures our findings are applicable and relevant to a wide range of real-world scenarios.

\subsubsection{Evaluation Metric}

To measure the effectiveness of \alias, we employ \textit{Compression Ratio} (CR) as the metric, which is widely used in the evaluation of existing compression methods for telemetry data~\cite{DBLP:conf/kbse/LiuZHHZL19,DBLP:conf/icse/LiZL024}.
The definition is given below:

\begin{equation*}
    CR=\frac{\mathrm{Original~File~Size}}{\mathrm{Compressed~File~Size}}
\end{equation*}

In each experiment, we run the same set of workloads, so the size of the original file remains constant.
With different compression approaches and configurations, the resulting compressed file may vary in size.
As the file size decreases, a higher level of compression is attained, indicating more effective compression performance.

\subsubsection{Baseline Methods}

Since we are the first to study the problem of trace compression in a live production scenario, there has not been any baseline methods/systems for comparison.
Note that \alias is orthogonal to existing trace sampling techniques, which compress traces via reducing the volume of data collected.
Thus, they cannot be directly compared to \alias.
In this case, we opt for general-purpose compression algorithms which can be used as out-of-the-box tools to compress traces.
Three prevalent and effective algorithms are selected, that is, gzip, bzip2, and lzma.
However, as they are not tailored for trace data, suggesting potential for further performance improvement.
Our goal is to illustrate the additional compression benefits that \alias can provide when applied in conjunction with these standard algorithms.

\begin{table*}[t]
    \centering
    \caption{Performance of Trace Compression on Open-source Cloud Systems}
    \label{tab:dynamic_compression_result}
    \centering
    \footnotesize
    \begin{NiceTabular}{C{1.89cm}|C{0.45cm} C{0.45cm}|C{0.45cm} C{0.45cm}|C{0.45cm} C{0.45cm}|C{0.45cm} C{0.45cm}|C{0.45cm} C{0.45cm}|C{0.45cm} C{0.45cm}|C{0.45cm} C{0.45cm}}
        \specialrule{0.35mm}{0em}{0em}
        \multirow{2}{*}{} & \multicolumn{2}{c}{\textbf{Train Ticket}} & \multicolumn{2}{c}{\textbf{gRPC}} & \multicolumn{2}{c}{\textbf{Kafka}} & \multicolumn{2}{c}{\textbf{Servlet}} & \multicolumn{2}{c}{\textbf{MySQL}} & \multicolumn{2}{c}{\textbf{Redis}} & \multicolumn{2}{c}{\textbf{MongoDB}}\\
        \cline{2-15}
        & \textbf{Size} & \textbf{CR} & \textbf{Size} & \textbf{CR} & \textbf{Size} & \textbf{CR} & \textbf{Size} & \textbf{CR} & \textbf{Size} & \textbf{CR} & \textbf{Size} & \textbf{CR} & \textbf{Size} & \textbf{CR}\\
        \specialrule{0.15mm}{0em}{0em}
        \specialrule{0.15mm}{.1em}{0em}
        Raw & 21.0 & 1 & 3.08 & 1 & 2.47 & 1 & 9.36 & 1 & 1.88 & 1 & 2.01  & 1 & 1.10 & 1 \\
        \rowcolor{grey} \alias & 5.19 & 4.05 & 0.58 & 5.31 & 0.627 & 3.94 & 1.45 & 6.46 & 0.30 & 6.27 & 0.33 & 6.09 & 0.21 & 5.24 \\
        \hdashline[2pt/1pt]
        gzip & 1.93 & 10.90 & 0.163 & 18.91 & 0.143 & 17.27 & 0.506 & 18.50 & 0.112 & 16.85 & 0.084 & 23.93 & 0.065 & 16.92 \\
        \rowcolor{lightgrey} \alias (gzip) & \textbf{1.29} & 16.26 & 0.140 & 22.81 & 0.133 & 18.57 & 0.396 & 23.64 & 0.091 & 20.61 & 0.066 & 30.45 & 0.051 & 21.57 \\
        \rowcolor{grey} improvement & 33.0\% & 1.49x & 17.0\% & 1.17x & 7.0\% & 1.08x & 21.7\% & 1.28x & 18.2\% & 1.22x & 21.4\% & 1.27x & 21.5\% & 1.27x \\
        \hdashline[2pt/1pt]
        bzip2 & 2.41 & 8.71 & 0.135 & 21.96 & 0.124 & 19.92 & 0.421 & 22.23 & 0.097 & 19.38 & 0.056 & 35.89 & 0.054 & 20.37 \\
        \rowcolor{lightgrey} \alias (bzip2) & 1.34 & 15.62 & \textbf{0.128} & 24.00 & 0.116 & 21.29 & 0.365 & 25.64 & 0.087 & 21.56 & 0.044 & 45.68 & 0.048 & 22.92 \\
        \rowcolor{grey} improvement & 30.3\% & 1.75x & 8.5\% & 1.10x & 6.5\% & 1.07x & 13.3\% & 1.15x  & 10.3\% & 1.12x & 21.4\% & 1.27x & 11.1\% & 1.13x \\
        \hdashline[2pt/1pt]
        lzma & 1.93 & 10.89 & 0.174 & 17.67 & 0.128 & 19.29 & 0.487 & 19.22 & 0.121 & 15.61 & 0.064 & 31.41 & 0.065 & 16.92 \\
        \rowcolor{lightgrey} \alias (lzma) & 1.55& 13.54  & 0.146 & 21.14 & 0.012 & 20.58 & 0.412 & 22.72 & 0.097 & 19.48 & 0.054 & 37.2 & 0.055 & 20 \\
        \rowcolor{grey} improvement & 35.7\% & 1.24x & 16.4\% & 1.20x & 9.6\% & 1.07x & 20.1\% & 1.18x & 19.9\% & 1.2x & 17.4\% & 1.19x & 22.7\% & 1.18x \\
        \specialrule{0.35mm}{0em}{0em}
    \end{NiceTabular}
\end{table*}

\subsection{Effectiveness of Trace Compression}
\label{sec:compression_effectiveness}

\subsubsection{Open-source Cloud Systems}

Table~\ref{tab:dynamic_compression_result} presents the compression performance when collecting traces of the microservices benchmark and cloud applications components.
For each system, we calculate the total size of traces collected, the size after compression, and the resultant compression ratios (CRs) when applying different compression algorithms.
We can see that \alias, as a standalone solution, can achieve CRs ranging from 3.94 to 6.46.
This demonstrates that \alias can remove more amount of redundant information than that shown by our preliminary study in Section~\ref{sec:redundancy_study}.
Traditional compression tools, i.e., gzip, bzip2, and lzma, reduce the file size with a combination of different techniques such as dictionary-based compression and Huffman coding.
Among them, bzip2 generally outperforms the others across most systems, with gzip having the least effectiveness.
In the Train Ticket benchmark, the tools demonstrate the least effective compression with a CR of roughly 10, while on the cloud application components, they deliver a better performance, attaining a comparable CR of around 20.

When working in conjunction with the general-purpose compression algorithms, \alias can provide additional performance gain.
In general, the improvement achieved by \alias when combined with bzip2 is less pronounced than when paired with other algorithms.
This can be attributed to its already superior compression capability, which may reduce the incremental benefits that \alias can offer.
In the case of the microservices benchmark, namely Train Ticket, \alias achieves a more significant performance improvement of 30\%$\sim$35\%.
However, the improvement is less substantial in cloud application components, with Apache Kafka demonstrating an enhancement of less than 10\%.
As mentioned in Section~\ref{sec:redundancy_study}, the traces generated by Kafka include the data from its message queues, rendering the attributes more random.

So far we can make an important observation: compared to application backend components, general-purpose compression algorithms are less effective for processing the traces from Train Ticket, where \alias can offer more substantial improvement.
Our careful investigation reveals the following important insight.
Based on zero-code instrumentation, the spans collected encapsulate many attributes related to network connectivity (as specified by OpenTelemetry semantic conventions), such as the hostname, IP address, and port of the peer server.
For instance, MongoDB captures details of the requests; Kafka producers log information about their consumers.
Such information provides a comprehensive view of the request's journey across the distributed system.
In production systems, the invocations among different services and components constitute a complex graph, with each node potentially connected to dozens or more instances.
Our experimental environment may not be able to accurately replicate the conditions of the production scenarios.
Consequently, the connectivity information tends to be relatively static, especially for application backend components that operate at infrastructure and platform layer.
In this case, both \alias and traditional algorithms can properly compress such information, reducing the performance gain that \alias can offer.

On the other hand, Train Ticket comprises tens of microservices, which can form a invocation graph with moderate complexity.
Additional, as a service-oriented application, the traces from Train Ticket contain more information related to business logic.
These two factors render the traces produced in Train Ticket more diverse, and the compressible information is more scattered.
Traditional compression algorithms are limited to exploiting redundant information within a short sliding window (e.g., 32KB in gzip's Deflate algorithm).
On the other hand, \alias utilizes \sname to continuously capture the redundancy patterns across spans in a \textit{global} manner, which can further reduce the redundancy.

\subsubsection{Production Trace Data}

We also evaluate \alias using production trace data collected from Alibaba. Compared to existing microservices benchmarks, this dataset represents the call graphs of a large-scale deployment of over 20,000 microservices in production clusters. The participating microservices can be categorized into two types: stateless services and stateful services. Stateless services operate independently of any stored state data, whereas stateful services, including databases and systems like Memcached, are required to maintain state information.
There are three types of communication paradigms between pairs of microservices: inter-process communication, remote invocation, and indirect communication. In addition to this diversity, the trace data also exhibit statistical characteristics typical of industry scale. For example, the size of microservice call graphs follows a heavy-tail distribution; there is a non-negligible fraction of hot-spot microservices; and the microservices can form highly dynamic call dependencies at runtime.
This real-world application allows us to examine \alias's efficacy in handling large-scale, complex data, which is crucial for understanding its potential in practical, production-level scenarios.

\begin{figure}[t]
    \centering
    \includegraphics[width=0.6\linewidth]{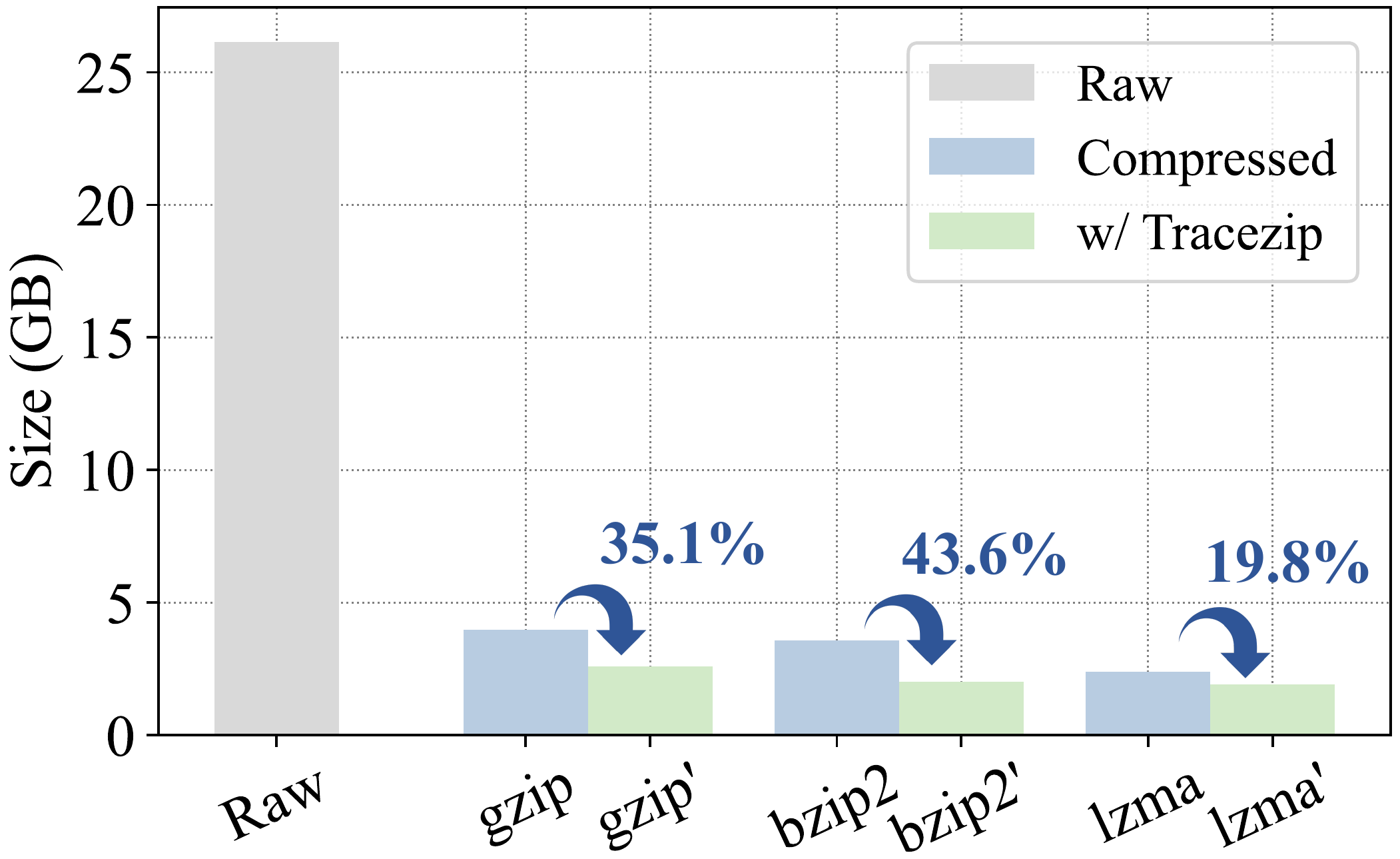}
    \caption{Compression on Alibaba Production Traces}
    \label{fig:alibaba_compression}
\end{figure}

Figure~\ref{fig:alibaba_compression} illustrates the evaluation results.
The raw size of the trace data used in our experiments is 26.15GB.
The CRs attained by gzip, bzip2, and lzma are 6.55, 7.30, and 8.52, respectively, reducing the data size to 3.99GB, 3.58GB, and 3.07GB.
These CRs are marginally lower than those recorded in the Train Ticket benchmark.
This observation aligns with our finding in Section~\ref{sec:compression_effectiveness}.
That is, relying predominantly on local information, traditional compression algorithms might find it challenging to efficiently compress data characterized by significant diversity and complexity.
This is where the strength of \alias becomes evident.
\alias, with its ability to identify global compression opportunities, enhances the compression performance by 35.1\%, 43.6\%, and 37.8\%, respectively.
For example, the combination of \alias and lzma achieves the optimal CR of 13.69, and compresses the data to a minimal size of 1.91GB.
This result underscores the benefits introduced by \alias, especially when dealing with large-scale and intricate trace data.

\subsection{Performance Overhead}

We examine the overhead of \alias from the perspectives of space complexity and computational efficiency.

\subsubsection{Space Complexity}

At the service side, \alias maintains two types of data structures to capture the redundancy among trace spans and perform compression, namely, a \sname along with its hashed paths and a map for dictionary-based compression.
To prevent impeding the normal execution of the service, it is imperative that they are constrained in size without excessive memory consumption.
To study the space complexity of \alias, we select a microservice in Alibaba trace data with diverse spans and calculate the cumulative size of the three data structures after the compression procedure.
A critical parameter influencing this size is the threshold $\psi$, which dictates the maximum number of distinct values for universal attributes.
An attribute having an exceeding number of values will be moved to the leaf node, becoming a local attribute.
A larger $\psi$ enables \alias to compress a broader spectrum of span fields, enhancing performance but at the cost of a more substantial SRT and mapping structure.
Conversely, a small $\psi$ compromises the effectiveness but with lower space overhead in return.

\begin{figure}
    \centering
    \includegraphics[width=0.66\linewidth]{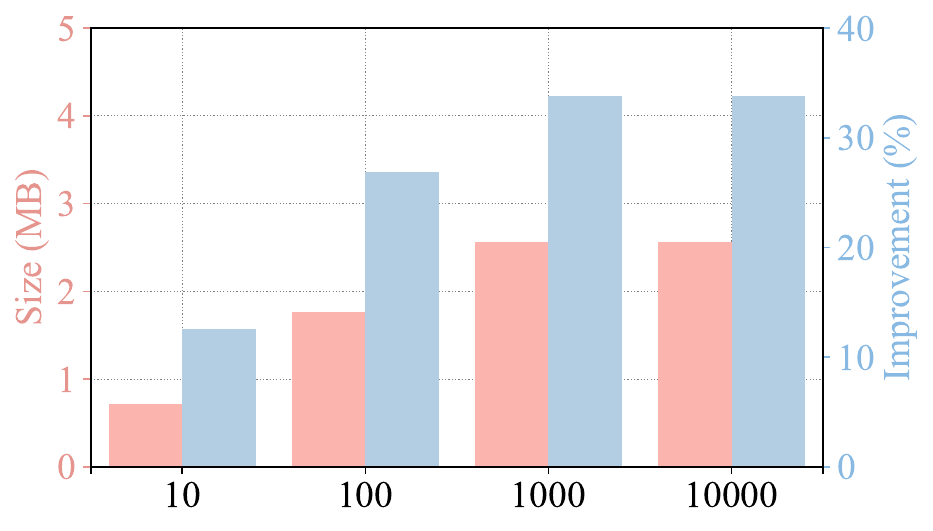}
    \caption{Performance with Different $\psi$}
    \label{fig:space_overhead}
\end{figure}

Figure~\ref{fig:space_overhead} illustrates the results, where we can see both the data structure size and compression improvement grow with a larger $\psi$.
In the case of $\psi$=1,000, \sname and map together take up only 2.56MB of memory, but the performance gain that \alias achieves is significant, i.e., 33.8\%.
This result underscores \alias's capacity to achieve substantial compression efficiency while maintaining a balanced memory footprint.
However, we notice that the quantities of distinct values that span attributes can have tend to polarize.
This can also be observed in Figure~\ref{fig:space_overhead}.
The performance plateaus even when $\psi$=10,000, meaning there is no attributes whose value size falls in the range of [1,000, 10,000].
Certain attributes (e.g., authentication tokens, DB queries, span ID) might possess a substantially larger set of values compared to others.
Consequently, their inclusion (when $\psi$ is too large) in the \sname could potentially bloat its size.
To address the variability in attribute value distribution and maintain manageable memory usage, we also set a cap on the size of the data structures, e.g., limiting it to 5MB.

\subsubsection{Computational Efficiency}

To ensure \alias can be seamlessly integrated with services, all operations are designed for optimal efficiency.
The time complexity of \alias's core operations is analyzed as follows.
The construction and restructuring of \sname operate with a time complexity of $\mathcal{O}(m)$, where $m$ is the number of span attributes.
In many scenarios, $m$ typically remains below 20.
Other operations, such as hashing, dictionary mapping, and path searching, all have a complexity of $\mathcal{O}(1)$.
Thus, the overall time complexity of \alias is \textit{linear}, making it highly efficient.

To evaluate the efficiency of \alias, we measure the trace collection throughput for the \textit{basic} microservice~\cite{trainticket} of the Train Ticket benchmark.
Specifically, we deployed the instrumented OpenTelemetry system within a container (configured with one core and 1GB of memory) to compress and relay the spans.
The throughput is calculated as the uncompressed size of spans divided by the time token to transmit the traces from the service to the backend.
In the most basic setting, referred to as \textit{Original}, the time is purely the period needed for data transmission and JSON serialization.
When compression techniques such as \alias and gzip are employed, we take into account the additional time required for data compression and decompression.
The results are present in Table~\ref{tab:throughput}.

\begin{table*}[h]
    \centering
    \caption{Performance of Throughput (MB/s) on Train Ticket}
    \label{tab:throughput}
    \centering
    \footnotesize
    \begin{NiceTabular}{C{2.5cm}|C{1cm}|C{1cm}|C{1cm}|C{1cm}}
        \specialrule{0.35mm}{0em}{0em}
        $\psi$ & \textbf{1} & \textbf{10} & \textbf{100} & \textbf{1,000} \\
        \specialrule{0.15mm}{0em}{0em}
        \specialrule{0.15mm}{.1em}{0em}
        Original & 13.98 & 13.57 & 13.78 & 14.05 \\
        \rowcolor{lightgrey} +\alias & 89.34 & 94.35 & \textbf{109.68} & 108.15 \\
        +gzip & 14.65 & 14.27 & 14.35 & 14.02 \\
        \rowcolor{lightgrey}+\alias (gzip) & 60.65 & 63.78 & 68.78 & 68.56 \\
        \specialrule{0.35mm}{0em}{0em}
    \end{NiceTabular}
\end{table*}

It can be seen that upon the integration of \alias, the throughput of trace collection is accelerated by nearly eight times (e.g., from 13.78MB/s to 109.68MB/s).
Another interesting observation is that gzip brings little performance gain to the throughput.
One important reason is that gzip compression is performed by the HTTP client library before the data is sent over the network.
Thus, gzip is applied after the data has been serialized into JSON.
As a time-consuming step, JSON serialization constitutes the performance bottleneck.
Moreover, gzip indiscriminately attempts to compress all information, including elements such as trace IDs, span IDs, and authentication tokens, which are inherently resistant to compression.
The (wasted) computational overhead of gzip compression and decompression thus offsets its benefits.
In contrast, \alias can accurately identify the incompressible attributes, i.e., the local fields, and bypass them.
Since \alias is applied before JSON serialization, it substantially reduces the volume of data that needs to be encoded.
Such a design not only improves the throughput, but also benefits the CPU usage.
Our experiments indicate that the CPU utilization of Tracezip is merely 20\%$\sim$40\% of that in the Original and +gzip settings.




\subsection{Threats to Validity}

When evaluating the performance and applicability of \alias, several potential threats to validity must be considered to ensure the robustness and generalizability of our findings.

\textbf{Internal validity}. One of the primary concerns regarding internal validity is the accuracy of our evaluation metrics and the potential biases in our experimental setup.
Real-world cloud services exhibit a vast array of complexities and variations, making it challenging to capture all possible scenarios within a single study.
To address this challenge, we carefully select a diverse set of microservices benchmarks and production trace data from Alibaba that we believe are representative of typical cloud service operations.
These benchmarks and dataset are chosen to reflect common patterns and behaviors observed in real-world applications, thus providing a meaningful context for evaluating \alias's performance.
Additionally, any configuration or tuning of \alias that is specific to these datasets might inadvertently favor our approach, potentially skewing the results.
To mitigate this, we ensure that the benchmarks and dataset are selected and configured independently of \alias's development process.

\textbf{External validity}. External validity pertains to the generalizability of our results to other settings or systems.
Our evaluation of \alias is specifically designed to address the diversity inherent in real-world cloud systems.
We implement and test \alias within the OpenTelemetry Collector framework and evaluate it across a range of cloud environments and backend components, including gRPC, Apache Kafka, MySQL, and others. 
These settings are carefully selected to reflect the variety of systems and technologies commonly used in cloud services, ensuring a comprehensive basis for assessing \alias's effectiveness.
By choosing such a diverse array of environments and applications, we aim to capture the broad spectrum of redundancy patterns and data characteristics found in typical cloud systems.
This approach helps to ensure that our findings are applicable to a wide range of real-world scenarios, demonstrating \alias's capability to perform effectively in diverse and dynamic cloud environments.
\section{Related Work}
\label{sec:related_work}

\subsection{Distributed Tracing Systems}

Distributed tracing offers a holistic, end-to-end perspective on data requests as they navigate the myriad services within a distributed application.
Pioneering works, such as Magpie~\cite{DBLP:conf/hotos/BarhamIMN03}, Whodunit~\cite{DBLP:conf/eurosys/ChandaCZ07}, X-Trace~\cite{DBLP:conf/nsdi/FonsecaPKSS07}, and Dapper~\cite{sigelman2010dapper}, have laid the foundation of tracing in distributed systems.
Recent studies take steps further to address specific challenges to pursue accurate and efficient tracing capabilities.
For example, Pivot Tracing~\cite{DBLP:conf/sosp/MaceRF15} and Canopy~\cite{DBLP:conf/sosp/KaldorMBGKOOSSV17} emphasize cross-application and cross-platform tracing to ensure seamless monitoring and diagnostics across diverse environments.
Panorama~\cite{DBLP:conf/osdi/HuangGLZD18} achieves more sophisticated observability by collecting inter-process and inter-thread error signals.
DeepFlow~\cite{DBLP:conf/sigcomm/ShenZXSLSZWY0XL23} is a non-intrusive distributed tracing framework for troubleshooting microservices.
It establishes a network-centric tracing plane with eBPF technique in the kernel. 
OpenTelemetry~\cite{opentelemetry} standardizes the APIs for telemetry data collection, instrumentation libraries, and semantic conventions.
It can be used with a broad variety of observability backends, such as Jaeger~\cite{jaeger}, Zipkin~\cite{zipkin}, and Prometheus~\cite{prometheus}.

\subsection{Trace-based System Management}

Besides the infrastructure for distributed tracing, trace data are extensively used in many system reliability management tasks.
The empirical study presented in~\cite{DBLP:journals/tse/ZhouPXSJLD21} demonstrates that the current industrial practices of microservice debugging can be enhanced by integrating appropriate tracing and visualization techniques.
MEPFL~\cite{DBLP:conf/sigsoft/Zhou0X0JLXH19} leverages system trace logs to perform latent error prediction and fault localization for microservice applications.
Some works employ traces to construct the dependency graph of microservices for root cause analysis.
This include machine learning-based approaches, such as random walk~\cite{DBLP:conf/sigmetrics/KimSS13}, PageRank~\cite{DBLP:conf/www/YuCCGHJWSL21}, hierarchical clustering and K-means~\cite{DBLP:journals/smr/YuHC23}, and spectral analysis~\cite{DBLP:conf/issre/ZhouZPYLLZZD23}, etc.
In recent years, trace-based fault diagnosis also resorts to deep learning-based approaches, e.g., 
\cite{DBLP:conf/asplos/GanZHCHPD19} combines CNN with LSTM to address the complexity of performance debugging.
TraceAnomaly~\cite{DBLP:conf/issre/ZhouZPYLLZZD23} uses deep Bayesian network to localize anomalous services in an unsupervised way based on trace representation learning.
Some work~\cite{DBLP:conf/sigsoft/YuCLCLZ23,DBLP:conf/kbse/ChenLSZWLYL21,DBLP:conf/icse/LeeYCSL23} utilizes a multi-modal approach, which integrates traces with logs~\cite{DBLP:journals/csur/HeHCYSL21,DBLP:journals/corr/abs-2107-05908,DBLP:conf/ccs/Du0ZS17} and metrics~\cite{DBLP:journals/corr/abs-2308-00393,DBLP:conf/icse/ChenL00LL22} to provide more comprehensive information about system status for microservice troubleshooting.
Traces also serve a crucial role in in analyzing system dependencies~\cite{DBLP:conf/cloud/LuoXLYXZDH021,DBLP:conf/nsdi/WuCP19}, critical paths~\cite{DBLP:conf/usenix/0002RRPSC22}, resource characterization~\cite{DBLP:conf/icpp/WangLWJCWDXHYZ22,DBLP:conf/kbse/LiuJGHCFYYL23}, and microservice architecture~\cite{DBLP:conf/usenix/HuyeSS23,DBLP:conf/sigsoft/0001ZZIGC22}.

\subsection{Trace Sampling and Compression}

In production environments, traces can carry comprehensive details, which, if not managed properly, can lead to significant overheads and potentially impact system performance.
To reduce overheads, the \textit{de facto} practice is to capture fewer traces.
Different from head-based sampling with random trace collection, tail-based solutions enable biased sampling towards more informative and uncommon traces.
Some learning-based approaches~\cite{DBLP:conf/IEEEcloud/ChenJSLZ24,DBLP:conf/icws/HuangCYCZ21,DBLP:conf/cloud/Las-CasasPAM19} have been proposed in this field.
Sieve~\cite{DBLP:conf/icws/HuangCYCZ21} employs Robust Random Cut Forest (RRCF) algorithm, a variant of Isolation Forest, to calculate an attention score for each trace, which is then used to determine its sampling probability.
Sifter~\cite{DBLP:conf/cloud/Las-CasasPAM19} captures edge-case traces by learning an unbiased, low-dimensional model to reconstruct the fixed-length sub-paths of traces.
A larger reconstruction loss indicates a higher sampling probability.
SampleHST~\cite{DBLP:conf/noms/GiasGSPOC23} and Perch utilize clustering techniques to divide traces into different groups, and sampling is then performed in each group.
STEAM~\cite{DBLP:conf/sigsoft/HeFLZ0LR023} preserves system observability by sampling mutually dissimilar traces.
It employs Graph Neural Networks (GNN) for trace representation, and requires human labeling to incorporate domain knowledge.
Hindsight~\cite{DBLP:conf/nsdi/ZhangXAVM23} introduces the idea of retroactive sampling, which combines the advantages of head-based and tail-based sampling.
Specifically, it allows the tracing for all requests at the service side, but reports trace data only after outlier symptoms are detected.

Different from prior research, \alias introduces an innovative methodology for mitigating the overhead associated with distributed tracing.
Rather than forecasting the significance of individual traces, \alias focuses on the compression of spans by leveraging their inherent redundancy.
Our approach can work with existing techniques complementarily to enhance the tracing performance.
It efficiently manages data volume while ensuring that the essential insights provided by trace data are preserved.

\section{Conclusion}
\label{sec:conclusion}

In this paper, we propose \alias, an online and scalable solution to mitigate the computational and storage overhead associated with distributed tracing.
Existing trace sampling techniques either sacrifices the accurate detection of edge cases or tracing scalability.
As an orthogonal approach, \alias reduces tracing overhead by compressing trace spans into a concise representation, substantially reducing the costs of trace transmission and storage.
This is achieved by removing the redundant data shared across multiple spans.
The full trace spans can then be seamlessly restored by retrieving the common data that are already delivered by previous spans.
At the service side, such redundancy is continuously encapsulated by Span Retrieval Tree (\sname), which will be synchronized with the backend to ensure consistent trace compression and uncompression.
To manage the complexity of \sname, \alias constantly restructures \sname into its optimal form and employs mapping technique to further reduces its size.
\alias also encompasses a differential update mechanism for efficiently synchronize \sname between services and the backend.
We have implemented \alias within OpenTelemetry Collector and evaluated it on open-source cloud services and production trace data.
Experimental results highlight its potential to pave the way for efficient system monitoring.


\section*{Acknowledgments}
The work described in this paper was supported by the National Natural Science Foundation of China (No. 62402536).
We extend our sincere gratitude to the anonymous reviewers for their insightful feedback.

\bibliographystyle{ACM-Reference-Format}
\bibliography{bibliography}

\end{document}